\documentclass[letterpaper, 10 pt, conference]{ieeeconf}  
\usepackage{amsmath, amssymb}
\usepackage{graphicx}
\usepackage{subfigure}
\usepackage[noadjust]{cite}
\usepackage{color}
\IEEEoverridecommandlockouts                              
\title{\LARGE \bf
Optimal control strategies for efficient energy harvesting from ambient vibrations
}

\author{Ashkan Haji Hosseinloo, Thanh Long Vu, and Konstantin Turitsyn
\thanks{This work was not supported by any organization}
\thanks{All the authors are with Department of Mechanical Engineering, 
Massachusetts Institute of Technology, 77 Massachusetts Ave, Cambridge MA
        {\tt\small \{ashkanhh, longvu, turitsyn\}@mit.edu}}%
}

\usepackage{graphicx}
\begin{document}

\maketitle
\thispagestyle{empty}
\pagestyle{empty}

\begin{abstract}
Ease of miniaturization and minimal maintenance are among the advantages for replacing conventional batteries with vibratory energy harvesters in a wide of range of disciplines and applications, from wireless communication sensors to medical implants. However, the current harvesters do not extract energy from the ambient vibrations in a very efficient and robust fashion, and hence, there need to be more optimal harvesting approaches. In this paper, we introduce a generic architecture for vibration energy harvesting and delineate the key challenges in the field. Then, we formulate an optimal control problem to maximize the harvested energy. Though possessing similar structure to that of the standard LQG problem, this optimal control problem is inherently different from the LQG problem and poses theoretical challenges to control community. As the first step, we simplify it to a tractable problem of optimizing  control gains for a linear system subjected to Gaussian white noise excitation, and show that this optimal problem has non-trivial optimal solutions in both time and frequency domains.  
\end{abstract}

\section{INTRODUCTION}
The problem of energy supply is one of the biggest issues in miniaturizing electronic devices. Advances in technology have reduced the power consumption in electronic devices, such as wireless sensors, data transmitters, and medical implants, to the point where ambient vibration has become a viable alternative to bulky traditional batteries \cite{hosseinloo2014fundamental}. In addition to scaling issues, recharging, replacing and disposing of batteries is usually cumbersome, costly, and could entail health-related and environmental complexities \cite{daqaq2014role}.


To further miniaturize electronic devices and to remedy the above-mentioned issues, energy harvesting has been investigated and considered as a scalable counterpart for batteries. Among many other sources, ambient vibration has captured attention in the last decade for its being universal and widely available.  Sources such as waves \cite{scruggs2009harvesting,wang2010piezoelectric}, bridge vibration \cite{elvin2006feasibility,galchev2011harvesting}, walking motion \cite{donelan2008biomechanical,rome2005generating,mitcheson2008energy}, and the movement of internal organs \cite{karami2012powering,dagdeviren2014conformal} are able to provide energy to a harvester. A typical vibratory energy harvester (VEH) consists of a vibrating host structure, a transducer, and an electrical load. A broad variety of different electromagnetic, electrostatic, piezoelectric, and magnetostrictive transduction mechanisms have been exploited in VEHs to convert the vibration energy of the host structure into useful electrical energy \cite{erturk2011broadband}.

The literature in inertial energy harvesting could be classified mainly into two categories: studies with emphasis on mechanical domain of the energy harvesters and studies with emphasis on energy harvesting circuitry (electrical domain). There are also some studies considering simplified models of the two domains at the same time, trying to maximize the harvested power (see for instance \cite{stephen2006energy,cammarano2010tuning}). The key challenges in vibration energy harvesting in both mechanical and electrical domains are achieving high efficiency under severe constraints (practical and inherent), robustness issues of the harvester, and multi-domain design complexities, to name a few. Practical constraints such as displacement constraints of the VEH or inherent transduction mechanism constraints impose upper-bound limit on maximum harvested energy. Broadband-spectrum or non-stationary excitations impose serious robustness issues on both mechanical oscillator and harvesting circuitry designs.

To overcome some of the aforesaid issues in the mechanical domain, researchers have used intentional nonlinearities, in particular mechanical bistability, in the hope to increase the energy flow to the system and make the system more robust to changes in the excitation. \cite{daqaq2014role} provides a comprehensive review and discussion for various types of nonlinearities studied in the literature. However, the system response and efficiency remains to be sensitive to the initial conditions (co-existing low-energy and high-energy orbits) \cite{erturk2009piezomagnetoelastic,mann2010investigations,stanton2010nonlinear,erturk2011broadband}, potential shape and acceleration intensity \cite{masana2011relative,daqaq2012intentional,litak2010magnetopiezoelastic,halvorsen2013fundamental,zhao2013stochastic,hosseinloo2015energy}, and nature of the excitation \cite{green2013energy}. Similar studies have been done in the harvesting circuitry design to increase the harvesting power available in the mechanical domain. \cite{szarka2012review,dicken2012power} provide recent reviews on different active and passive harvesting circuitry designs for optimal power conditioning and extraction. 

Although there is still much room for improvement in the power harvesting circuitry, there is larger room for improvement in the mechanical domain of the VEHs. The latter is a necessary step for effective and sufficient power delivery to the electrical domain. The linear and the current nonlinear VEHs cannot pump energy from the excitation sources to the harvesting circuitry in a very effective and robust way. The authors believe the powerful machinery developed in the controls contexts could substantially improve robust design and analysis of the VEHs in electrical and particularly, mechanical domains. 

To this end, we present in this paper a general electromechanical architecture of energy harvesting system and discuss several key challenges in details. Then, we will present a reduced model of a VEH with capacitive (piezoelectric) harvesting circuitry with additional passive control forces in both mechanical and electrical domains. On top of this model, we formulate the optimal control problems for maximizing the energy harvested. This problem generally involves maximizing a quadratic cost function over all the dissipative controls of a time-invariant linear system perturbed by Gaussian white noise. Though its structure is similar to that of the standard LQG problem, this control problem is not a LQG problem due to the dissipative constraint on the control and the missing of control penalty in the cost function. It is therefore challenging to solve this problem. In this paper, we outline possible ways to simplify and solve this problem. Our simulations show that there are many opportunities for improvement in the control design to maximize the harvested energy.

The main contributions of this paper include:
\begin{itemize}
\item [1)] We mathematically model the energy harvesting architecture;
\item [2)] We explicitly formulate the optimal control problem to maximize the energy harvested. This new optimal control problem is challenging and
requires new tools from control systems theory; and 
\item [3)] We simplify this optimal control problem to tractable problems and outline possible ways in both time and frequency domains to obtain the optimal control design.
\end{itemize}

The paper is organized as follows. In Section \ref{sec.Architecture}, the general architecture of the vibration energy harvesting system is introduced. Section 
\ref{sec.Modelling} presents the detailed mathematical model of a simplified energy harvesting architecture. In Section \ref{sec.Problem}, the general optimal control problem is formulated for maximizing energy harvesting, and then is simplified to tractable optimal control problems. Section \ref{sec.Simulations} numerically illustrates 
the optimal control obtained by direct calculation and simulations on frequency domain. Finally, in Section \ref{sec.Conclusions} we conclude the paper and discuss possible ways in the future to improve the model and the proposed control techniques, as well as suggesting several aspects where control expertise is necessary to leverage the energy harvesting industry.

\section{Energy Harvesting Architecture}
\label{sec.Architecture}
\begin{figure}
 \centering
 \includegraphics[width = \linewidth]{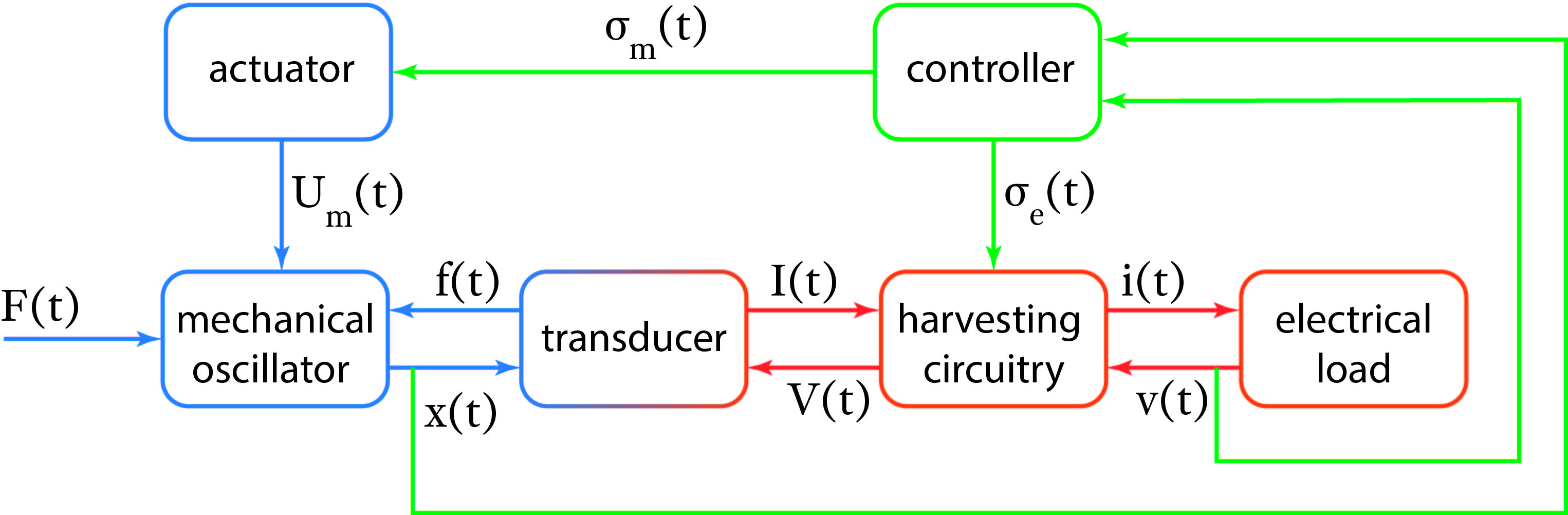}
 \caption{A generic energy harvesting architecture.$\mathrm{F}(t)$ denotes the excitation force of the energy source. $\mathrm{x}(t)$, and $\mathrm{f}(t)$ represent the mechanical domain states, and the electromechanical coupling force, respectively.$\mathrm{V}(t)$ and $\mathrm{v}(t)$ denote electrical voltage states, and $\mathrm{I}(t)$, and $\mathrm{i}(t)$ show the input current to their respective blocks. $\sigma_m(t)$ and $\sigma_e(t)$ represent control signals to the mechanical and electrical actuators, and $\mathrm{U_m}(t)$ denotes the force applied on the mechanical domain by the actuator.}
 \label{Fig1}
\end{figure}

Figure \ref{Fig1} depicts a generic architecture for vibration energy harvesting. The architecture is composed of four different sections: a mechanical domain which is usually a mechanical oscillator, an electrical domain which is usually a harvesting circuitry, a transduction mechanism which couples the mechanical and electrical domains (usually a piezoelectric, electromagnetic or electrostatic mechanism), and a control part which is usually not introduced or analyzed in detail in the vibration energy harvesting context. For better readability and for the sake of clarity these domains and their related signals are drawn with different colors in Fig. \ref{Fig1}.

The energy flows to the mechanical domain from the energy source e.g. vibration of a bridge or wave motion, and then through the electromechanical coupling to the electrical domain, and is then harvested through the harvesting circuitry. The controller based on its logic derives mechanical and electrical actuators to increase the energy flow to the mechanical oscillator, to the electrical domain, and ultimately to the electrical load, in an active, passive or a hybrid fashion. The electrical actuator is absorbed in the harvesting circuitry block in Fig.\ref{Fig1}. If properly designed, the controller can improve the robustness and efficiency of the harvester. Next, the architecture is realized with a simple model and the problem of maximizing the harvested energy is formulated as a control problem.

\section{Mathematical modelling}
\label{sec.Modelling}

\begin{figure}
 \centering
 \includegraphics[width = \linewidth]{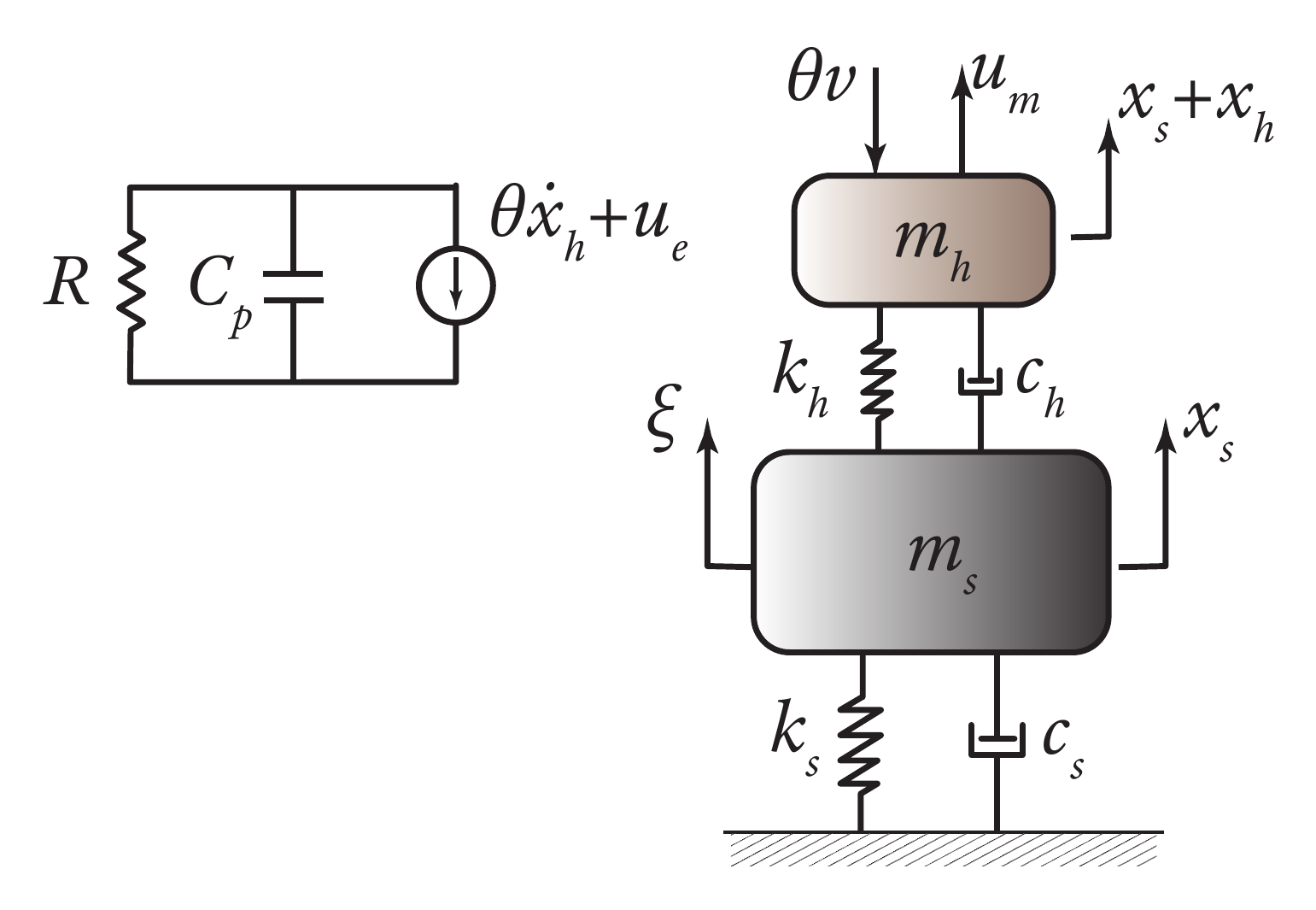}
 \caption{a simplified model of a piezoelectric energy harvester with mechanical ($u_m$) and electrical ($u_e$) control inputs, mounted on a sdof structure subjected to arbitrary excitation force $\xi(t)$ }
 \label{Fig2}
\end{figure}

In this section we present a simple lumped model of an energy harvester with one mechanical and one electrical degrees of freedom mounted on top of a structure that is also modeled as a single-degree-of-freedom (sdof) system (representing the first mode of a real structure such as a bridge or a building) that is excited by an arbitrary force (see Fig.\ref{Fig2}).

The first structural mode usually carries most of the kinetic energy of the structure (the dominant mode) and hence the structure acts as a low-pass filter between the excitation input and the harvester. Also, in practice, the harvester mass is usually negligible compared to the structure mass ($m_h\ll m_s$); consequently, the dynamics of the structure is not affected by the dynamics of the harvester. Thus, dynamics of the structure could be effectively described by its first mode as,
\begin{equation}
\ddot{x}_s+2\zeta_s\lambda \dot{x}_s +\lambda^2 x_s=\xi(t),
\label{Eq1}
\end{equation}
where, $x_s$, $\zeta_s$, and $\lambda$ are non-dimensional displacement of the structure, modal damping ratio of the structure, and ratio of the natural frequency of the structure to that of the harvester, respectively. $\xi(t)$ is exogenous excitation acting on the structure. 

The harvester oscillator is also modelled as a sdof system whose dynamics are driven by the base excitation via the structure, the control force, and the electromechanical coupling force. The dynamics of the harvester oscillator is governed by,
\begin{equation}
\ddot{x}_h+2\zeta_h \dot{x}_h + x_h +\kappa^2 v=-\ddot{x}_s +u_m(t),
\label{Eq2}
\end{equation}
where, $x_h$, $\zeta_h$, and $\kappa^2$ are non-dimensional displacement of the harvester relative to the structure, modal damping ratio of the harvester, and non-dimensional electromechanical coupling, respectively. $u_m(t)$ is the non-dimensional control force, and $v$ is non-dimensional electrical voltage whose dynamics are described by,
\begin{equation}
\dot{v} + \alpha v = \dot{x}_h +u_e(t).
\label{Eq3}
\end{equation}

In Eq.(\ref{Eq3}), $u_e(t)$ is the non-dimensional electrical current and $\alpha$ is the ratio between the mechanical and electrical time constants.

In the above equations, the displacements, voltage and time are non-dimensionalized by the quantities $l_c$, $\theta l_c/C_p$, and $1/\omega_h$, where $l_c$ is a scaling length, $C_p$ is the capacitance of the piezoelectric element, $\theta$ is the linear electromechanical coupling coefficient, and $\omega_h$ is the undamped nominal natural frequency of the harvester and is defined as $\sqrt{k_h/m_h}$. Note that overdot represents differentiation with respect to the non-dimensional time.

Also, in Eqs. \ref{Eq1}-\ref{Eq3}, the parameters are defined as,

\begin{align}
\zeta_s&=\frac{c_s}{2\sqrt{k_s m_s}}    &  \omega_s&=\sqrt{\frac{k_s}{m_s}}     &  \kappa^2&=\frac{\theta^2}{C_p k_h}\nonumber\\
\zeta_h&=\frac{c_h}{2\sqrt{k_h m_h}}    &  \lambda&=\frac{\omega_s}{\omega_h}   &  \alpha&=\frac{1}{RC_p\omega_h},\nonumber\\
\label{Eq4}
\end{align}
where, $m$, $k$ and $c$ represent mass, linear stiffness and damping, respectively and the subscripts $h$ and $s$ refer to the harvester and the structure as depicted in Fig.\ref{Fig2}. $R$ represents the electrical load resistance.



Now the objective here is to maximize the average non-dimensional output power (or energy) harvested through the resistance load,
\begin{equation}
\mathrm{maximize} \quad \overline{P}=\frac{1}{T} \int_{0}^{T} v(t)^2 \mathrm{d}t \footnote{The nondimensional $\overline{P}$ power is related to the dimensional power $\overline{P}^{\mathrm{dim}}$ by the relation $\overline{P}^{\mathrm{dim}}=(m_h \omega_h^3 l_c^2 \alpha \kappa^2)\overline{P}$ }.
\label{Eq5}
\end{equation}

{\color{black} Here we have assumed that the controllers  are passive; otherwise for an active system, the net energy injection to the system should be reflected in Eq.\ref{Eq5}. Passivity constraint for the mechanical and electrical controllers could be written as,

\begin{align}
\label{Eq6a}
u_m(t)\dot{x}_h \leq \dot{V}_m\left(x(t)\right),\quad u_e(t)v \leq \dot{V}_e(x(t))
\end{align}
where $V_m$ and $V_e$ are differentiable state-dependent storage (potential) functions in the mechanical and electrical control systems, respectively. The controller is called lossless (conservative) if the equality holds, otherwise the controller is said to be strictly passive (dissipative).
}



Also, in practice, we usually have constraints on the magnitude (maximum and/or minimum) of the control inputs as well. Moreover, due to volume constraints or to prevent mechanical failure, there is usually constraints on the maximum displacement of the harvester $x_h$. This constraint is not necessarily to be satisfied by the control forces, and in fact, it is usually satisfied by nonlinear dynamics of the system (which is not considered here) e.g. by stoppers or container walls.  

In the next section, we cast the problem into an optimal control problem and with some simplifying assumptions, will optimize the control forces $u_m$ and $u_e$.

\section{Optimal Control Problem Formulation}
\label{sec.Problem}




In this section, we particularly formulate the optimal control problem for maximizing the harvested energy. 
Define the state vector $x=[x_1 ... x_5]^T$ where $x_1=x_s, x_2=\dot{x}_s, x_3=x_h, x_4=\dot{x}_h, x_5=v.$ Then, the overall system \eqref{Eq1}-\eqref{Eq3}
is expressed as
\begin{align*}
\dot{x}_1 &=x_2 \\
\dot{x}_2 &=\xi(t) - 2\zeta_s\lambda x_2 - \lambda^2 x_1 \\
\dot{x}_3 &= x_4 \\
\dot{x}_4 &= u_m(t) - \xi(t) + 2\zeta_s\lambda x_2 + \lambda^2 x_1 - 2\zeta_h x_4 -  x_3 -\kappa^2x_5\\
\dot{x}_5 &= u_e(t) +   x_4 - \alpha x_5
\end{align*}
Equivalently, this set of equations can be written in the compact form:
\begin{align}
\label{eq.System}
\dot{x} =A x + B_u u +B_{\xi} \xi(t), 
\end{align}
where
\begin{align*}
A &=\left[%
\begin{array}{ccccc} 
0 & 1 & 0 & 0 & 0 \\
- \lambda^2 & - 2\zeta_s\lambda & 0 & 0 & 0 \\
0 & 0 & 0 & 1 & 0 \\
\lambda^2 & 2\zeta_s\lambda & - 1 & - 2\zeta_h & -\kappa^2 \\
0 & 0 & 0 & 1 & -\alpha \\
\end{array}%
\right], \\
B_u &=\left[%
\begin{array}{cc}
0 & 0 \\
0 & 0 \\
0 & 0 \\
1 & 0 \\
0 & 1 \\
\end{array}%
\right], B_{\xi} = [0 \;1 \;0 \;-1 \;0]^T
\end{align*} 
Here, $u= [u_m(t) \; u_e(t)]^T$ is the control input. Assume that the excitation force $\xi(t)$
is modeled as a Gaussian zero-mean white noise with variance $W \ge 0.$ 

Since matrix $A$ is naturally stable and the control is required to be {\color{black}passive}, the closed-loop system is stable. Therefore, the processes in the system are ergodic. Hence,
the cost function in \eqref{Eq5} can be rewritten as $J=\lim_{t\to\infty} \textbf{E} [v^T(t)v(t)].$ Formally, our objective is to design the control law $u$ to maximize the
energy harvested  described by the cost function 
\begin{align}
\label{eq.CostFunction}
J=   \lim_{t\to\infty}  \textbf{E} [v^T(t)v(t)] 
\end{align}
where $v= C x, C=[0 \;0 \;0\;0 \;1].$ Hence, we have the following optimal control problem:

\begin{itemize}
\item [\textbf{(P1)}:] \textbf{Optimal control for energy harvesting:} \emph{Given the system \eqref{eq.System}, design
the optimal and {\color{black}passive} control $u=[u_m(t)  \;  u_e(t)]^T$ to maximize the cost function $J$ defined in \eqref{eq.CostFunction}
subject to the constraints on the state $x$.}
\end{itemize}

We note that though the considered system \eqref{eq.System} is a linear system with Gaussian white noise and the cost function \eqref{eq.CostFunction} is quadratic, the general optimal control problem $\textbf{(P1)}$ is not a typical LQG
problem because there is no strict penalty on the input in the cost function and the control here need to be {\color{black}passive}. Also, while the LQG control minimizes the cost function with little control effort, the problem $\textbf{(P1)}$ tries to maximize the output $ \textbf{E} [v^T(t)v(t)]$ over all the {\color{black}passive} controls.

Therefore, the optimal control problem $\textbf{(P1)}$ is a new control problem. It is challenging and needs tools from control and optimization community. Having its similar structure with the LQG problem, we may use the similar technique
to solve this general problem. Another possible way to solve this problem is to utilize the Pontryagin's maximum principle. This approach may be complicated when we use higher order models for the structure and harvester.

In the following, we simplify the optimal control problem $\textbf{(P1)}$ and outline the way to obtain the optimal control $u,$
even for higher-order model.
Basically, due to the noise $\xi(t)$ there is no perfect prediction for the state $x$ of the system and the control law $u$ should be in the form of a filter-based control. In this paper, we assume that we have perfect prediction for the state $x$ and the control law $u$ is just a function of $x.$ In addition, to satisfy the {\color{black}passivity} requirement of the controllers, we will choose conservative (lossless) control inputs of the following simple form:
{\color{black}
\begin{align}
u_m(t) = -K_m x_h, \qquad u_e(t) = -K_e \dot{v}
\label{Eq9}
\end{align}
}
where $K_m>0$ and $K_e>-1$. This type of controller guarantees that the control forces are {\color{black}passive (conservative in this case) and also practically implementable. In fact, $u_m$ in Eq.\eqref{Eq9} represent a spring with nondimensional spring constant $K_m$ connecting the harvester to the structure, and $u_e$ represents a capacitor in parallel $(K_e>0)$ or in series $(-1<K_e<0)$ with the inherent piezoelectric capacitor}.\footnote{They could be implemented even when $K_m$ and $K_e$ are variable with variable spring and capacitor.} For simplicity, we do not consider the constraints on the state $x$. 

Finally, we simplify the considered optimal control problem $\textbf{(P1)}$ into the following problem:

\begin{itemize}
\item [\textbf{(P2)}:] \textbf{Optimal control gains:} \emph{Find the optimum values for the control gains  $K_m$ and $K_e$
to maximize the cost function $J:$
\begin{align}
J^* = \max_{K_m>0, K_e >-1}   \lim_{t\to\infty} \textbf{E} [x^T(t)C^TCx(t)],
\end{align}
where the dynamics of $x(t)$ is described by
\begin{align}
\dot{x}&=A_Kx + B_{\xi}\xi, \\
A_K&= \left[%
\begin{array}{ccccc} 
0 & 1 & 0 & 0 & 0 \\
- \lambda^2 & - 2\zeta_s\lambda & 0 & 0 & 0 \\
0 & 0 & 0 & 1 & 0 \\
\lambda^2 & 2\zeta_s\lambda & - (1+K_m) & - 2\zeta_h & -\kappa^2 \\
0 & 0 & 0 & \dfrac{1}{1+K_e} & -\dfrac{\alpha}{1+K_e} \\
\end{array}%
\right]. 
\end{align}}
\end{itemize}

To solve this problem, we can directly calculate $J$ using the Controllability Gramian of the system \eqref{eq.System}:
\begin{align}
J= \textbf{Tr}(C^TCP),
\end{align} 
where $P$ is positive definite solution of the Lyapunov equation $A_K^TP+PA_K+ B_{\xi}WB_{\xi}^T=0.$ Therefore,
we have the following optimization
\begin{align}
\label{eq.optimalControl}
J^* &= \max \textbf{Tr}(C^TCP), \\
s.t. \quad &K_m>0,K_e>-1, \nonumber\\
& P>0, \nonumber\\
&A_K^TP+PA_K+ B_{\xi}WB_{\xi}^T=0. \nonumber
\end{align}
By solving this optimization using some Optimization ToolBoxs, we can obtain the optimum values for the control gains $K_m$ and $K_e.$ We note that
the constraint $A_K^TP+PA_K+ B_{\xi}WB_{\xi}^T=0$ leads to a \textbf{stable} closed-loop system 
with the optimum control $u=[-K_m^*x_h \;\; -K_e^*\dot{v}].$

\section{Frequency-domain approach}
\label{sec.Simulations}

An alternative approach to the optimization problem formulated above to find the optimum gains, is a brute-force optimization in the frequency domain. The latter is substantially easier for the problem at hand in \textbf{(P2)}, mainly because of the simple form of the excitation in the frequency domain (white noise Gaussian) and simple form of the control force and current considered in Eq. \eqref{Eq9}.

In view of the Parseval's theorem, instead of maximizing the average power (Eq. \eqref{Eq5}) in the time domain one could maximize it in the frequency domain, that is,

\begin{equation}
\mathrm{maximize} \quad \lim_{T\to\infty}\overline{P}=\frac{1}{T} \int_{0}^{T} v(t)^2 \mathrm{d}t=\int_{-\infty}^{+\infty} |V(\omega)|^2 \mathrm{d}\omega,
\label{Eq15}
\end{equation}
where $V(\omega)$ is the Fourier transform of the the voltage $v(t)$. Since the governing dynamic Eqs. \eqref{Eq1}-\eqref{Eq3} are linear, it is easy to solve for $V(\omega)$ in terms of the system parameters and Fourier transform of the input excitation $\Xi(\omega)$ (which is simply the transfer function from from the input $\xi(t)$ to the output $v(t)$ in the frequency domain). It could be easily shown that,

\begin{equation}
V(\omega)=\frac{A(\omega)B(\omega)D(\omega)}{1-C(\omega)D(\omega)}\Xi(\omega),
\end{equation}
where, 

\begin{align}
A(\omega)&= \frac{1}{\lambda^2-\omega^2+2\zeta_s\lambda\omega i}    \nonumber\\
B(\omega)&= \frac{\omega^2-K_m}{1+K_m-\omega^2+2\zeta_h\omega i}      \nonumber\\
C(\omega)&= \frac{-\kappa^2}{1+K_m-\omega^2+2\zeta_h\omega i}     \nonumber\\
D(\omega)&= \frac{\omega i}{\alpha-K_e+\omega i}.
\end{align}

\begin{figure}
 \centering
 \includegraphics[width = 1\linewidth]{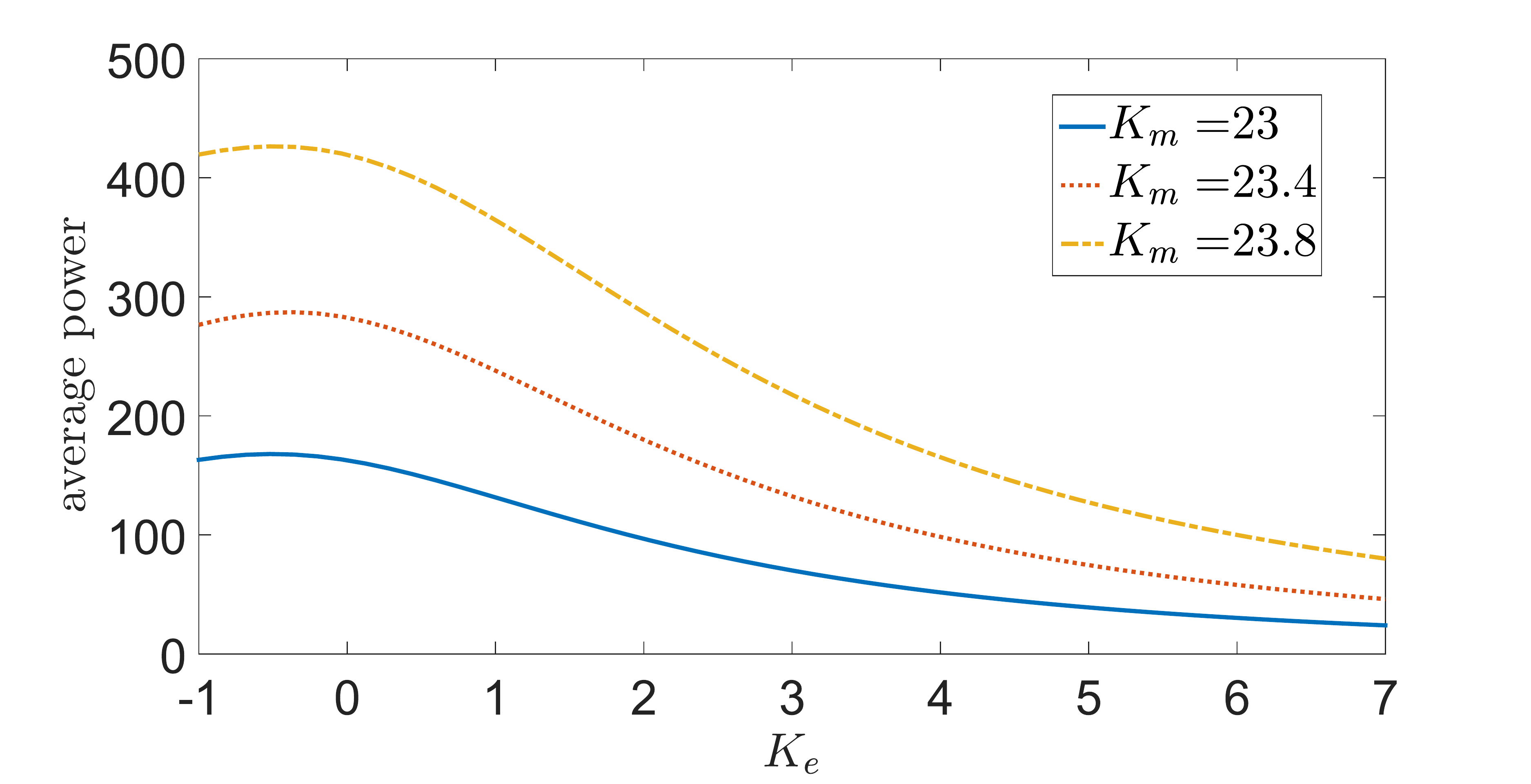}
 \caption{Scaled harvested energy as a function of $K_e$ for different values of $K_m=$ 0.1, 0.3, and 0.5. The parameters are set as $\lambda$=5, $\zeta_s$=0.01, $\zeta_h$=0.01, $\kappa$=0.6, and $\alpha$=10 }
 \label{Fig3}
\end{figure}

\begin{figure}
 \centering
 \includegraphics[width = 1\linewidth]{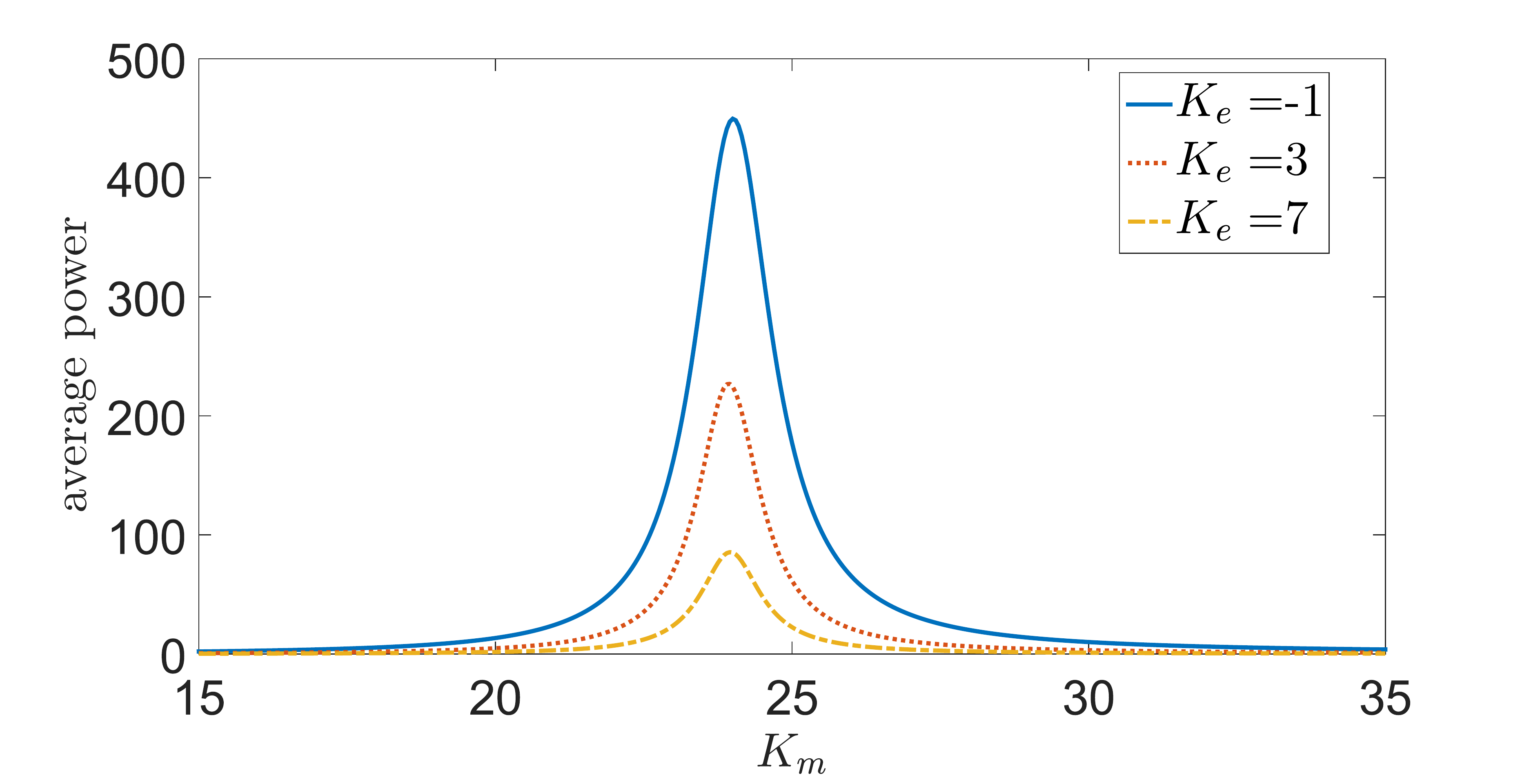}
 \caption{Scaled harvested energy as a function of $K_m$ for different values of $K_e=$ 900, 925, and 950. The parameters are set as $\lambda$=5, $\zeta_s$=0.01, $\zeta_h$=0.01, $\kappa$=0.6, and $\alpha$=10  }
 \label{Fig4}
\end{figure}

To find the optimum gains, one can calculate the integral in Eq. \eqref{Eq15} for a range of $K_m$ and $K_e$ gains and look for the optimum gains. Figure \ref{Fig3} depicts the average harvested power as a function of $K_e$ gain for different values of gain $K_m$. This figure shows the optimum gain for $K_e$ for given system parameters and $K_m$. Figure \ref{Fig4} shows dependence of the average power on $K_m$ for different values of the gain $K_e$. This figure also reveals optimum gain for $K_m$ for a set of given parameters. A nondimentional power spectral density of $|\Xi(\omega)|^2=1$ is used for the simulations.

The results here prove that there are non-trivial and optimum solutions to the optimal control problems defined and formulated in section \ref{sec.Problem}. The frequency-domain analysis breaks down when more complicated control laws are used, or more complex constrains are applied on the system or system and/or controller become nonlinear or when the excitation is not easily expressed in the frequency domain. However, the powerful machinery developed in the controls could still be applied and optimize the harvester designs.  

%

\section{Conclusions and path forward}
\label{sec.Conclusions}

This paper was dedicated to bring the fast-growing area of energy harvesting to the attention of control expertise. For this purpose, 
we have sketched a general architecture of energy harvesting from  ambient vibrations, highlighted the key challenges 
in this area, and showed how to come up with the optimal energy harvesting. To facilitate rigorous approaches tackling these challenges, we presented a simple
yet practically generic and efficient mathematical model of this architecture and pointed out the control options to leverage the energy harvesting process.
On top of this mathematical model, we explicitly formulated the optimal control problem to maximize the harvested energy. It should be noted that
though this optimal control problem possess similar structure to that of the standard LQG problem, it is inherently different from the LQG problem in twofolds: (i) there is no penalty on the control input in the cost function, which serves to maximize the given output function of the system (i.e. the energy harvested) (ii) the optimal control itself needs to be of a particular characteristics {\color{black}e.g. passivity}. As the first step to resolve this challenging problem, we simplified it to a tractable problem of optimizing the control gains for linear system subjected to Gaussian white noise excitation, and outlined possible ways in both time and frequency domains to come up with the optimal control design. Our numerical simulations showed that there are many opportunities to maximize the energy harvesting based on solving these optimal control problems. 

We envision several aspects where control expertise is indispensable to push the current framework to the practical level. First, the new optimal control problem introduced in this paper, even in its simple form, is challenging and requires sophisticated new tools from the optimal control theory. Second, a higher-order model should be developed to capture the complicated dynamics of the system in practice, while the constraints on the states and controls should be considered. Finally, we need to investigate the robustness of the controlled VEH performance when there are uncertainty in the system model and when more complicated broadband-spectrum or non-stationary excitations are considered.




\bibliography{CDCManuscript.bbl}






\end{document}